\documentclass[twocolumn,showpacs,preprintnumbers,prl,aps,amssymb,superscriptaddress]{revtex4}
\usepackage{graphicx}
\usepackage{amsmath}
\usepackage{dcolumn}
\usepackage{bm}
\begin{document}

\newcommand{\q}{{\bf{\it q}}}


\title{ Heat Transport as a Probe of Electron Scattering by Spin Fluctuations:\\
the Case of Antiferromagnetic CeRhIn$_5$ }


\author{Johnpierre~Paglione}
\altaffiliation[Current address: ] {Department of Physics, University of California, San Diego, La Jolla, CA.}
\affiliation{Department of Physics, University of Toronto, Toronto, Ontario, Canada}

\author{M.A.~Tanatar}
\altaffiliation[Permanent address: ] {Inst. Surface Chemistry, N.A.S. Ukraine, Kyiv, Ukraine.}
\affiliation{Department of Physics, University of Toronto, Toronto, Ontario, Canada}

\author{D.G.~Hawthorn}
\affiliation{Department of Physics, University of Toronto, Toronto, Ontario, Canada}

\author{R.W.~Hill}
\affiliation{Department of Physics, University of Toronto, Toronto, Ontario, Canada}

\author{F.~Ronning}
\affiliation{Department of Physics, University of Toronto, Toronto, Ontario, Canada}

\author{M.~Sutherland}
\affiliation{Department of Physics, University of Toronto, Toronto, Ontario, Canada}

\author{Louis Taillefer}
\email{Louis.Taillefer@USherbrooke.ca}
\affiliation{Department of Physics, University of Toronto, Toronto, Ontario, Canada}
\affiliation{D\'epartement de physique et Regroupement qu\'eb\'ecois sur les mat\'eriaux de pointe, Universit\'e de Sherbrooke, Sherbrooke, Qu\'ebec, Canada}
\affiliation{Canadian Institute for Advanced Research, Toronto, Ontario, Canada}

\author{C.~Petrovic}
\affiliation{Department of Physics, Brookhaven National Laboratory, Upton, New York 11973}

\author{P.C.~Canfield}
\affiliation{Ames Laboratory and Department of Physics and Astronomy, Iowa State University, Ames, Iowa 50011}

\date{\today}


\begin{abstract}

Heat and charge conduction were measured in the heavy-fermion metal CeRhIn$_5$, an antiferromagnet with $T_N=3.8$~K. The thermal resistivity is found to be proportional to the magnetic entropy, revealing that spin fluctuations are as effective in scattering electrons as they are in disordering local moments. The electrical resistivity, governed by a ${\bf q}^2$ weighting of fluctuations, increases monotonically with temperature. In contrast, the {\it difference} between thermal and electrical resistivities, characterized by a $\omega^2$ weighting, peaks sharply at $T_N$ and eventually goes to zero at a temperature $T^{\star} \simeq 8$~K.  ~$T^{\star}$ thus emerges as a measure of the characteristic energy of magnetic fluctuations.

\end{abstract}

\maketitle


The impact of a magnetic instability on the behaviour of electrons can be studied in a controlled way by tuning a magnetic transition to absolute zero temperature. At this quantum critical point (QCP), critical fluctuations are known to cause Fermi-liquid theory to fail and, in some materials, superconductivity to appear \cite{Stewart}. Several theories have been proposed to account for electronic transport in systems close to an antiferromagnetic QCP, invoking for example ``hot spots'' on the Fermi surface in a spin-density-wave model, local quantum criticality, or composite fermions \cite{Coleman}.

To test such theories, it is important to have information on the momentum and energy dependence of the magnetic fluctuation spectrum and its impact on electron transport. Part of this information can be obtained from inelastic neutron scattering. In this Letter, we present a complementary approach which relies on heat transport to shed light on the nature of magnetic scattering. The idea is to compare heat and charge conductivity in a given material, exploiting the fact that fluctuations affect the two differently, in a way which depends on energy and momentum. This technique can be applied where neutron scattering is difficult, {\it e.g.} with very small samples or in high magnetic fields. A typical case would involve the study of materials with a field-tuned QCP ({\it e.g.} Sr$_3$Ru$_2$O$_7$ \cite{Grigera}, YbRh$_2$Si$_2$ \cite{Custers} or CeCoIn$_5$ \cite{Paglione}).


Here we present a detailed study of heat and charge transport in a well-characterized material where spin fluctuations dominate the scattering of electrons. The material is CeRhIn$_5$, a metal with an incommensurate antiferromagnetic ground state below $T_N=3.8$~K \cite{Bao}. 
We report two observations: 
1) the thermal resistivity tracks the magnetic entropy perfectly, revealing that spin fluctuations are just as effective in scattering electrons as they are in disordering moments; 
2) the difference between thermal and electrical resistivities provides a direct measure of the characteristic energy of the fluctuation spectrum.


Single crystals of CeRhIn$_5$ were grown by the self-flux method \cite{Cedomir}. Their high quality is confirmed by a remarkably low residual resistivity, $\rho_0 = 0.037~\mu\Omega$~cm. Samples were prepared into rectangular parallelepipeds with typical dimensions $\sim 4 \times 0.1 \times 0.05$~mm. Electrical contacts for standard four-wire measurements were made with indium solder, resulting in low contact resistances ($\sim 5$~m$\Omega$). The electrical resistivity $\rho$ was measured with an AC resistance bridge and the thermal conductivity $\kappa$ was measured by a standard one-heater/two-thermometer technique. All measurements were performed using the same four contacts, with currents applied in the basal plane of the tetragonal crystal structure.


\begin{figure}
 \centering
 \includegraphics[totalheight=2.7in]{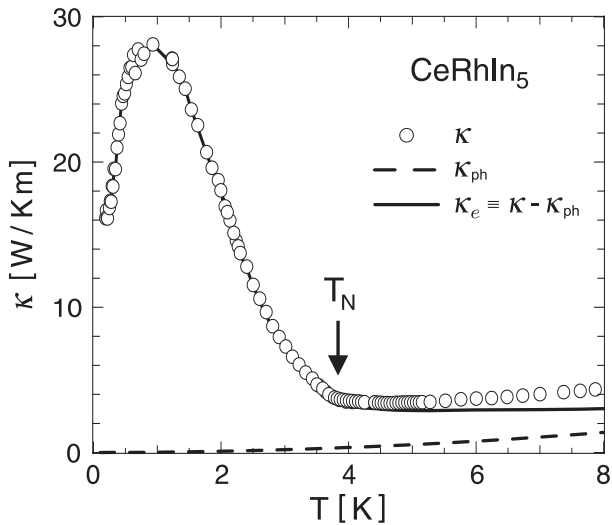}
 \caption{\label{fig:kappa} Thermal conductivity of CeRhIn$_5$, for a current in the basal plane (open circles). The dashed line is the estimated phonon conductivity ($\kappa_{ph}$), and the solid line the electronic contribution ($\kappa_e \equiv \kappa - \kappa_{ph}$). }
\end{figure}

The temperature dependence of the thermal conductivity of CeRhIn$_5$ is shown in Fig.~\ref{fig:kappa}. As in any metal, $\kappa$ is the sum of an electronic ($\kappa_e$) and a phononic ($\kappa_{ph}$) contribution. The conductivity of phonons in CeRhIn$_5$ was estimated by measuring $\kappa(T)$ in the isostructural and closely related material CeCoIn$_5$, in which 2\% La impurities were introduced to ensure that elastic impurity scattering dominates over the intrinsic inelastic scattering. In such a case, $\kappa_e$ may be obtained from the Wiedemann-Franz law: $\kappa_e(T) = L_0 T / \rho(T)$, where $L_0 \equiv \frac{\pi^2}{3}\left(\frac{k_B}{e}\right)^{2} = 2.44 \times 10^{8}$ ~W$~\Omega$~K$^{-2}$, and $\rho$ is the measured electrical resistivity. As a result, the phonon contribution is given by $\kappa_{ph}(T) \simeq \kappa(T) - L_0 T/\rho(T)$. A fit to the temperature dependence of $\kappa_{ph}(T)$ thus obtained, assumed to be the same for CeRhIn$_5$ \cite{phonon}, is shown as a dashed line in Fig.~\ref{fig:kappa}. The resulting electronic conductivity of CeRhIn$_5$, defined as $\kappa_e \equiv \kappa - \kappa_{ph}$ \cite{magnon}, deviates from the measured $\kappa$ only slightly (by approximately 10\% at $T_N$). From here onwards we focus on $\kappa_e$. 

Upon cooling, $\kappa_e$ increases dramatically below the onset of antiferromagnetic order at $T_N$, such that $\kappa_e/T$ grows by a factor of 60. This is due to the freezing out of magnetic fluctuations upon entering the ordered state. In order to explore this connection in detail, we compare the thermal resistivity $w_e(T) \equiv L_0 T / \kappa_e(T)$ (in units of $\rho$) to two other quantities: the magnetic entropy $S_{\rm mag}(T)$ and the electrical resistivity $\rho(T)$.

\begin{figure}
 \centering
 \includegraphics[totalheight=2.7in]{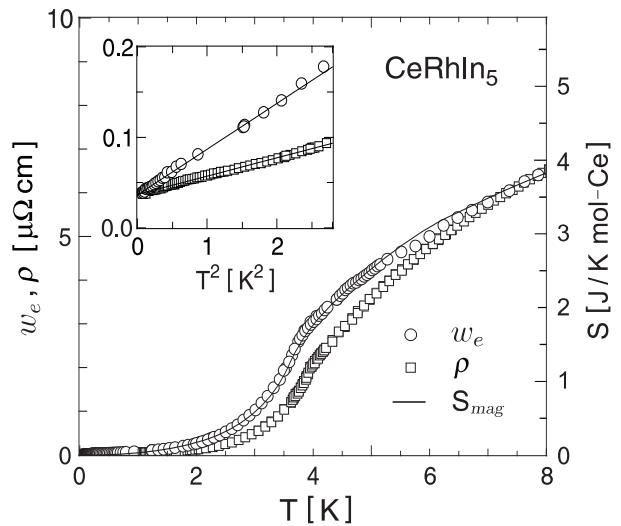}
 \caption{\label{fig:rho} Electronic thermal resistivity $w_e$ (circles) and electrical resistivity $\rho$ (squares), compared to the magnetic entropy $S_{\rm mag}$ (line) obtained from published specific heat data \cite{Hegger}. 
Inset: low-temperature data as a function of $T^2$. Lines are linear fits.}

\end{figure}

In Fig.~\ref{fig:rho}, $w_e(T)$ is seen to perfectly track $S_{\rm mag}(T)$, calculated from specific heat measurements by Hegger {\it et al.} \cite{Hegger}, over a wide range of temperature ($0<T<2T_N$). Such a relation, $w_e(T)/S_{\rm mag}(T)={\rm const.}=1.68~\frac{\mu\Omega~{\rm cm}}{{\rm J/K~mol-Ce}}$ , has to our knowledge never been discovered before. Fisher and Langer pointed out that the same spin-spin correlation function enters in the calculation of both the magnetic energy and the relaxation time associated with scattering of electrons by spin fluctuations, so that the temperature derivative of the resistivity $d\rho/dT$ should vary as the magnetic specific heat $C_{\rm mag}$ near $T_N$ \cite{Fisher}. This predicted correlation was roughly confirmed in measurements on the antiferromagnet PrB$_6$, for example, where a sharp peak was observed in both $d\rho/dT$ and $C_{\rm mag}(T)$ at $T_N=6.9$~K \cite{PrB6}.


The same approximate correlation ($d\rho/dT \propto C_{\rm mag}$) was pointed out in the case of CeRhIn$_5$ by Bao {\it et al.} who showed it to originate from the magnetic correlation function, measured with neutron scattering \cite{Bao}. Fig.~\ref{fig:rho} reveals that in CeRhIn$_5$ the best correlation is in fact between {\it scattering rate} and {\it entropy} ($w_e \propto S_{\rm mag}$), rather than $d\rho/dT \propto C_{\rm mag}$. Moreover, it holds much better for heat transport than for charge transport, presumably because charge conductivity involves a stronger angular weighting of fluctuations (in favour of high-{\bf q}) than heat conductivity, while entropy involves none.

\begin{figure}
 \centering
 \includegraphics[totalheight=2.7in]{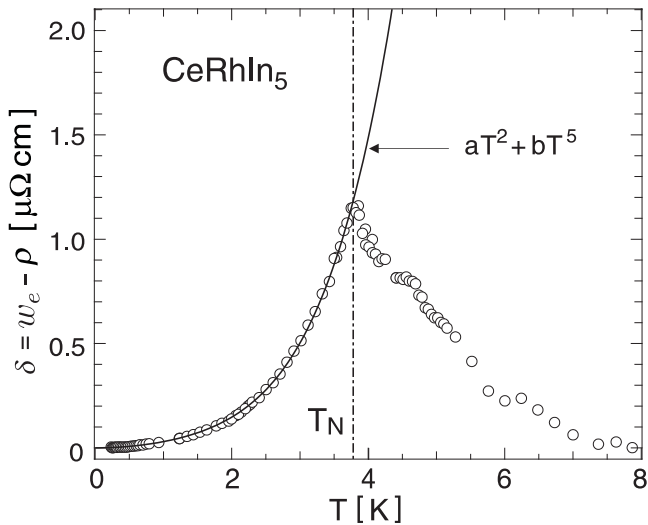}
 \caption{\label{fig:difference}
Difference between thermal ($w_e \equiv L_0 T / \kappa_e$) and electrical ($\rho$) resistivities: $\delta(T) \equiv w_e(T) - \rho(T)$. The vertical dash-dotted line marks the N\'eel temperature $T_N$. Note how abruptly the onset of static antiferromagnetic order cuts off the growth in $\delta(T)$ with decreasing temperature. Note also that $\delta(T)$ vanishes above $T \simeq$~8~K, revealing that temperature has by then exceeded the characteristic energy of magnetic fluctuations. 
The solid line is a fit to $aT^2 + bT^5$ below $T_N$.}
\end{figure}

\begin{figure}
 \centering
 \includegraphics[totalheight=2.7in]{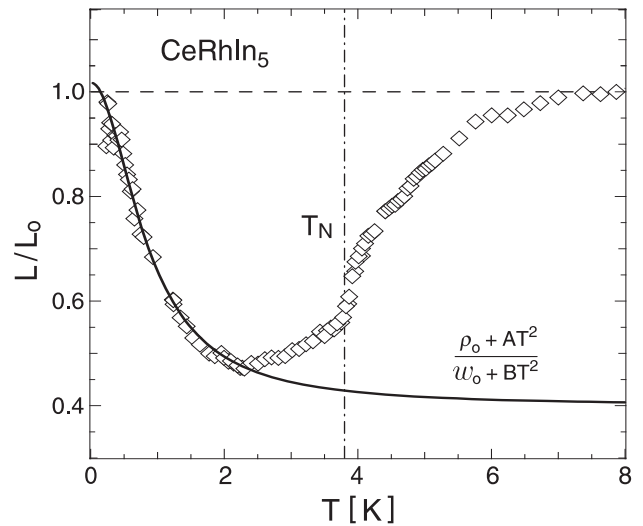}
 \caption{\label{fig:Lorenz} 
 Temperature dependence of normalized Lorenz ratio, $L(T)/L_0 \equiv \kappa_e / L_0 \sigma T = \rho(T) / w_e(T)$. The solid line is a fit to the Fermi-liquid expression, $L/L_0 = (\rho_0 + A T^2) / (w_0 + B T^2)$. }
\end{figure}

We compare the two resistivities in detail, both through their difference, defined as $\delta(T) \equiv w_e(T) - \rho(T)$ and shown in Fig.~\ref{fig:difference}, and through their ratio, defined as the normalized Lorenz ratio $L(T)/L_0 \equiv \rho(T) / w_e(T)$ and shown in Fig.~\ref{fig:Lorenz}. A scattering event degrades a charge current ($j_{\rho}$) and a heat current ($j_w$) by different amounts \cite{Schriempf}:
\begin{align}
\Delta j_{\rho} ~\simeq &~-~ \frac{k_F}{m^{\star}}~ e ~ ( 1 - {\rm cos}\theta ) \\
\Delta j_{w} ~\simeq &~-~ \frac{k_F}{m^{\star}}~ [~ (E - \mu) ~ ( 1 - {\rm cos}\theta ) ~+ ~\hbar \omega~ {\rm cos}\theta~ ]
\end{align}
where $e$ is the electron charge and $\mu$ the Fermi energy. The electron has initial velocity $\hbar k_F / m^{\star}$ and energy $E$, and sees its direction deflected by an angle $\theta$ and its energy changed by an amount $\hbar \omega$. For elastic scattering ($\hbar \omega = 0$), both currents are degraded in the same way, namely by a change in momentum direction. In this regime, one obtains the Wiedemann-Franz law, as indeed confirmed in CeRhIn$_5$ at $T \to 0$~: $w_e(T) = \rho(T)$ (Fig.~\ref{fig:rho}), or $L(T) = L_0$ (Fig.~\ref{fig:Lorenz}).

However, for inelastic scattering (finite $\hbar \omega$), the two terms in Eq.~2 lead to two contributions to the thermal resistivity, such that $w_e = w_{\rm hor} + w_{\rm ver}$ \cite{Kaiser}, but only the first type of scattering process enters in $\rho$, so that $\rho(T) = w_{\rm hor}(T)$ and $\delta(T) = w_{\rm ver}(T)$. These two scattering processes are sometimes referred to as ``horizontal'' and ``vertical'' processes, resulting, respectively, from changes in the direction of the electron wavevector and changes in the electron energy \cite{Schriempf,Kaiser}.

Both terms in $w_e$ are weighted integrals over ${\bf q}$ and $\omega$ of the fluctuation spectrum: $w_{\rm hor}$ is weighted by ${\bf q}^2$, while $w_{\rm ver}$ is weighted by $\omega^2$. Comparing the two gives access to the ${\bf q}$ and $\omega$ dependence of magnetic scattering. Calculations of electrons scattering off fluctuating local moments show that the effect of vertical processes is greatly reduced (eventually to zero) as the temperature increases above $\omega^{\star}$, the characteristic temperature of the spin fluctuations, since these fluctuations then have insufficient energy to scatter electrons through the thermal layer (width of the Fermi function) \cite{Kaiser}. This effect is well-known in the case of phonon scattering where $L \to L_0$ (and hence $\delta(T) \to 0$) when $k_B T > \hbar \omega_D$, where $\omega_D$ is the Debye frequency, the characteristic frequency of lattice fluctuations. The conclusion is that $\delta(T)$ can be used to determine $\omega^{\star}$ while $\rho(T)$ by itself typically cannot.

In this light, let us now examine the behaviour of $\delta(T)$ and $L(T)$ in CeRhIn$_5$. In Fig.~\ref{fig:difference}, $\delta(T)$ is seen to exhibit two key features: (1) it vanishes for $T > 8$~K and (2) it drops abruptly below $T_N$. The first feature reveals the sharp contrast between horizontal and vertical scattering processes, which respectively cause $w_{\rm hor}(T)$ to increase steadily with increasing temperature, but $w_{\rm ver}(T)$ to {\it decrease} (beyond $T_N$). In analogy with phonon scattering, we use the fact that $\delta \to 0$ at $T > 8$~K to claim that the characteristic fluctuation energy in CeRhIn$_5$ is of the order of 8 K. Actual calculations, along the lines of those by Kaiser \cite{Kaiser} but with an appropriate fluctuation spectrum, are needed to be more specific, but note that {\bf q}-dependent magnetic correlations observed by neutron scattering in CeRhIn$_5$ do have a characteristic energy less than 1.7 meV (18~K) and they develop below 7~K \cite{Bao}.

The second, rather dramatic, feature of Fig.~\ref{fig:difference} is the fact that the rise in $\delta(T)$ with decreasing $T$ is interrupted abruptly by the onset of static antiferromagnetic order at $T_N$. Indeed, immediately below $T_N$, $\delta(T)$ drops rapidly with a dependence which is well described by a smooth power law of $\delta(T) = w_{\rm ver}(T) = aT^2 + bT^5$ at all temperatures below $T_N$. Let us look at both terms in turn.

As shown in the inset of Fig.~\ref{fig:rho}, a $T^2$ dependence is observed in CeRhIn$_5$ for both resistivities, below $\sim$~1.5~K: $\rho=\rho_0+AT^2$ and $w_e=w_0+BT^2$, with $A=0.021~\mu\Omega$~cm/K$^2$ and $B=0.057~\mu\Omega$~cm/K$^2$. The magnitude of $A$ is quite small compared to other heavy-fermion metals such as CeCoIn$_5$ \cite{Paglione}, but in fact the ratio of $A$ to the electronic specific heat coefficient $\gamma=56$~mJ/K$^2$/mol~Ce \cite{Cornelius} yields a Kadowaki-Woods ratio ($A/\gamma^2=6.7 \times 10^{-6}~\mu\Omega$~cm~K$^2$~mol$^2$/mJ$^2$) that lies on the universal line for heavy-fermion compounds \cite{KW}.

The fact that $B > A$ reflects the importance of vertical processes and low-{\bf q} scattering. In this so-called Fermi-liquid regime, we therefore have $\delta(T) \sim T^2$ and $L/L_0 = (\rho_0 + A T^2) / (w_e + B T^2)$ (solid line in Fig.~\ref{fig:Lorenz}), so that the {\it inelastic} Lorenz ratio is constant: $L_{\rm in}(T) \equiv (\rho(T) - \rho_0)/(w_e(T) - w_0) = A / B = 0.4$. Quantitatively, the precise value of $L_{\rm in}$ is sensitive to the angular distribution of scattering over the Fermi surface \cite{L_ee}. In practice, a ratio $A/B \simeq 0.4$-$0.6$ is characteristic of most metals \cite{Wagner}, from elemental Ni, where $A/B \simeq 0.4$ \cite{White_Ni}, to the heavy-fermion compound UPt$_3$, where $A/B \simeq 0.65$ \cite{Lussier}. A calculation based on a two-band model of $s$-electrons from a spherical Fermi surface scattered by $d$-electrons from a cylindrical surface, applied to the transition metal Re, gave $L_{\rm in}=0.4$ \cite{Schriempf}, the same value observed in our experiments. The application of such a model to CeRhIn$_5$ is justified given the presence of light, spherical 3D pockets and heavy, quasi-2D cylindrical sheets in the Fermi surface \cite{dHvA}. Finally, note that electron scattering off localized spin fluctuations also yields a $T^2$ dependence and a ratio $A/B$ in the range 0.3-0.6, depending on the angular distribution of scattering \cite{Schriempf,Kaiser}.

The $T^5$ term in $\delta(T)$ below $T_N$ is a distinctive property of vertical scattering in the ordered state. It survives all the way up to $T_N$, but then abruptly goes away beyond that temperature (see Fig.~\ref{fig:difference}). The impact of broken symmetry is dramatic, first and foremost because of a suppression of spin fluctuations caused by a change in the fluctuation spectrum. [This is perhaps due to the opening of a gap, suggested by an activated dependence of specific heat \cite{Cornelius}, although no obvious exponential dependence is seen in $\delta(T)$.] That suppression is reflected in $\rho(T)$ as well, but it leads to a slightly different temperature dependence in the case of horizontal scattering: $\rho(T) - \rho_0 = AT^2 + cT^6$. Because this contrast contains information on the nature of magnetic fluctuations in the ordered state, it would be interesting to correlate the apparent differences between $\rho(T)$ and $\delta(T)$ with another measure of spin fluctuations: the drop in the sub-lattice magnetization $M(T)$, measured with neutron diffraction \cite{Bao} and $^{115}$In nuclear quadrupole resonance \cite{Curro}.


In conclusion, we have shown that the dual measurement of heat and charge transport in a metal with magnetic scattering can be used to probe the {\bf q} and $\omega$ dependence of spin fluctuations and their effect on electron scattering. Our study on the test material CeRhIn$_5$ reveals a number of interesting features:
(1) the thermal resistivity is directly proportional to the magnetic entropy;
(2) the difference between heat and charge transport vanishes above 8 K
-- a result which can be used to obtain the characteristic energy of the fluctuation spectrum;
(3) the inelastic Lorenz ratio is equal to 0.4 at low temperature -- a direct measure of the angular distribution of scattering;
(4) the onset of antiferromagnetic order yields a temperature dependence in the thermal resistivity due to vertical scattering ($\omega^2$ weighting) that is different than that of the electrical resistivity due to horizontal scattering  ({\bf q}$^2$ weighting).
Detailed calculations based on the known fluctuation spectrum of CeRhIn$_5$ would be very useful in further exploring this information.

{\it Note added.} -- In analogy with the use of heat transport by {\it electrons} to probe magnetic excitations in this antiferromagnetic metal, heat transport by {\it phonons} has recently been used as a ``spectroscopy'' of magnons in the antiferromagnetic insulator Nd$_2$CuO$_4$ \cite{Li}.


This work was supported by the Canadian Institute for Advanced Research and a Canada Research Chair (L. T.) and funded by NSERC. The authors gratefully acknowledge useful discussions with Y.~B.~Kim and M.~Smith, and M. F. Hundley and J. D. Thompson for specific heat data. Work at Brookhaven was supported by the Division of Materials Sciences, Office of Basic Energy Sciences, U.S. Department of Energy under contract No.
DE-AC02-98CH10886.



\begin{references}
\bibitem{phonon} A Ce-based (rather than La-based) compound was used to estimate $\kappa_{ph}$ in order to avoid severely altering the environment of electrons, which are the main source of scattering for phonons at the temperatures under study. The use of CeCoIn$_5$ is adequate since its thermal conductivity is similar to that of CeRhIn$_5$ above $T_N$ (below $T_N$, the phonon contribution to $\kappa$ is negligible).
\bibitem{magnon} Magnon contributions to $\kappa$ are known to be severely limited by scattering from charge carriers, both in lightly doped insulators \cite{mag_ladder} and magnetic metals \cite{mag_alloy}, and are thus negligible.
\end{references}
\end{document}